\documentclass[APA,STIX1COL,12pt]{WileyNJD-v2}

\usepackage{graphicx}
\usepackage{float}  
\usepackage{subcaption}
\usepackage{textcomp}
\usepackage{tabularx}
\usepackage{booktabs}     
\usepackage{siunitx}      
\usepackage[section]{placeins} 
\usepackage{adjustbox}
\usepackage{caption}
\usepackage{bm}
\usepackage{setspace}

\articletype{Article Type}%

\received{}
\revised{}
\accepted{}

\begin{document}

\title{Spatio-temporal Shared-Field Modeling of Beluga and Bowhead Whale Sightings Using a Joint Marked Log–Gaussian Cox Process} \protect
\author[1]{Mauli Pant*}
\author[2]{Linda Fernandez}
\author[3]{Indranil Sahoo}
\authormark{Pant \textsc{et al.}}

\address[1]{\orgdiv{Integrative Life Sciences Doctoral Program, Center for Integrative Life Sciences Education}, \orgname{Virginia Commonwealth University}, \orgaddress{\state{Virginia}, \country{USA}}}
\address[2]{\orgdiv{VCU School of Life Sciences and Sustainability, VCU Department of Economics}, \orgname{Virginia Commonwealth University}, \orgaddress{\state{Virginia}, \country{USA}}}
\address[3]{\orgdiv{Department of Statistical Sciences and Operations Research}, \orgname{Virginia Commonwealth University}, \orgaddress{\state{Virginia}, \country{USA}}}

\corres{*Mauli Pant \email{pantm2@vcu.edu}}


\abstract[Abstract]{%
We analyze a decade of aerial survey whale sighting data (2010–2019) to model the spatio-temporal distributions and group sizes of beluga (\textit{Delphinapterus leucas}) and bowhead (\textit{Balaena mysticetus}) whales in the United States Arctic. To jointly model these species, we develop a multi-species Log-Gaussian Cox Process (LGCP) in which species specific intensity surfaces are linked through a shared latent spatial Gaussian field. This structure allows the model to capture broad spatial patterns common to both species while still accommodating species level responses to environmental covariates and seasonal variation. The latent field is represented using the Stochastic Partial Differential Equation (SPDE) approach with an anisotropic Matérn covariance, implemented on an ocean constrained triangulated mesh so that spatial dependence aligns with marine geography. Whale group size is incorporated through a marked point process extension with species specific negative binomial marks, allowing occurrence and group sizes to be jointly analyzed within a unified framework. Inference is carried out using the Integrated Nested Laplace Approximation (INLA), enabling efficient model fitting over a decade of survey effort. The results highlight persistent multi-species hotspots and distinct environmental associations for each species, demonstrating the value of shared field LGCPs for joint species distribution modeling in data sparse and heterogeneous survey settings.%
}


\maketitle

\section{Introduction}\label{sec:intro}

The Arctic Northwest Passage (NWP) has experienced a steady rise in vessel traffic in recent years, with further growth projected over the coming decades \citep{CMTS2019}. The navigation season coincides with peak presence of bowhead (\textit{Balaena mysticetus}) and beluga (\textit{Delphinapterus leucas}) whales in the Chukchi and Beaufort Seas. Bowheads overwinter in the northern Bering Sea, migrate through the Chukchi in spring, and spend summers in the eastern Bering Sea and Beaufort region  \citep{MarineMa18online}. Belugas are Arctic endemic and exhibit pronounced seasonal movements across coastal corridors and channels \citep{Citta2020}. Increased shipping therefore elevates risks to whales via disturbance, ship strike, and noise exposure. Understanding the spatial distribution of whale groups is thus critical for ecological conservation and for maritime safety planning where shipping routes intersect core habitats.

Prior ecological studies have emphasized the influence of environmental and spatial covariates on Arctic cetacean distributions. For belugas in the Chukchi and Beaufort Seas, \citet{Hauser2017} found only limited direct effects of sea ice on habitat selection after accounting for spatial structure, while \citet{doi:10.1126/sciadv.abn2422} reported that temperature related variables, including sea surface temperature (SST), are likely to become increasingly important determinants of Arctic cetacean habitat suitability under future warming. Spatial predictors such as distance to coast and bathymetric depth consistently emerge as influential. For instance, \citet{clarke2020distribution} documented offshore shifts in bowhead whale distribution during low ice years, reflecting behavioral flexibility under changing conditions, and \citet{Clark2018Arctic} reported pronounced seasonal and depth related patterns in long term aerial survey records from the Beaufort Sea. At broader scales, telemetry analyses reveal seasonal redistribution linked to topographic features and foraging opportunities \citep{Citta2018}.

Building on this literature, we analyze a decade (2010--2019) of United States Arctic aerial survey data to characterize whale group distributions and their environmental associations. Extending beyond ensemble species distribution models such as Random Forest and Generalized Additive Models (GAMs) \citep{doi:10.1126/sciadv.abn2422} and resource‐selection approaches applied to belugas in the region \citep{Hauser2017}, as well as descriptive distribution analyses from long‐term aerial surveys \citep{clarke2020distribution}, we adopt a spatio‐temporal Log‐Gaussian Cox Process (LGCP) to model whale sightings as realizations of a continuous point process over space and time. In this framework, observed group locations arise from an inhomogeneous Poisson process whose intensity varies with environmental covariates and a latent Gaussian random field.

A key feature of our formulation is that beluga and bowhead whale sightings are modeled jointly using species specific LGCP intensity surfaces, so that each species can respond differently to environmental covariates, monthly structure, and behavior. Spatial dependence is represented by two latent Gaussian fields, one for each species, defined on a common SPDE mesh and governed by the same anisotropic Mat\'ern prior. This implies shared assumptions about spatial range, smoothness, and anisotropy, while still allowing distinct spatial patterns for belugas and bowheads. Temporal variation is captured with  monthly effects for July–October and year effects for 2010–2019, treated as random effects to reflect seasonal migration and inter annual variability in survey effort and habitat. Group size is modeled conditionally on location with a species specific negative binomial mark model, so that occurrence intensity and group size distributions are linked through the latent fields but can vary independently across space, time, and behavior. By operating directly on the point pattern of sightings rather than on binned count data, the marked LGCP framework provides coherent probabilistic uncertainty quantification for both whale occurrence and expected group sizes, while explicitly accounting for residual spatial structure.

For computationally efficient inference over large, complex marine domains, we use the stochastic partial differential equation (SPDE) approach with Integrated Nested Laplace Approximation (INLA), implemented in \texttt{inlabru}/\texttt{R-INLA} \citep{lindgren2011explicit}. The SPDE representation defines approximate Mat\'ern Gaussian random fields on a finite element mesh, allowing flexible latent spatial effects, seasonal structure, and covariate responses while scaling efficiently to a decade of survey data. In doing so, our study contributes to Arctic whale spatial ecology in four main ways.
First, we assemble and harmonize 2010–2019 Aerial Surveys of Arctic Marine Mammals (ASAMM) sightings with key environmental covariates (SST, distance to coast) to build a high resolution spatio temporal dataset for the Beaufort Archipelago. Second, we fit a marked LGCP that jointly models (i) the spatial intensity of whale group occurrences and (ii) the conditional distribution of group size, enabling ecological inference on both where the groups occur and how large they tend to be. Third, we compare species specific models to a joint, species grouped SPDE formulation that shares spatial hyperparameters across species, testing whether common spatial scales govern co occurrence while retaining species specific fields. Finally, we generate monthly predictive maps with uncertainty and evaluate goodness of fit using information criteria and point process diagnostics, providing a reproducible framework for risk aware navigation assessment in the United States Arctic.



\section{Aerial Surveys of Arctic Marine Mammals (ASAMM) whale sighting data}

We analyze a spatio-temporal dataset of whale group sightings obtained from systematic aerial surveys conducted by the Aerial Surveys of Arctic Marine Mammals (ASAMM) program across the eastern Chukchi and western Beaufort Seas from 2010-2019 \citep{Clarke2010, Clarke2011, Clarke2012, Clarke2013, Clarke2014, Clarke2015, Clarke2016, Clarke2017, Clarke2018, Clarke2019}. Surveys followed standardized protocols, with observers flying predetermined transects at consistent altitudes and speeds to ensure comparable detectability across years. At each sighting, observers recorded species identity, group size, behavior, environmental conditions, and the aircraft’s GPS location.
\begin{figure}[!ht]
    \centering
    \footnotesize \includegraphics[width=0.9\linewidth]{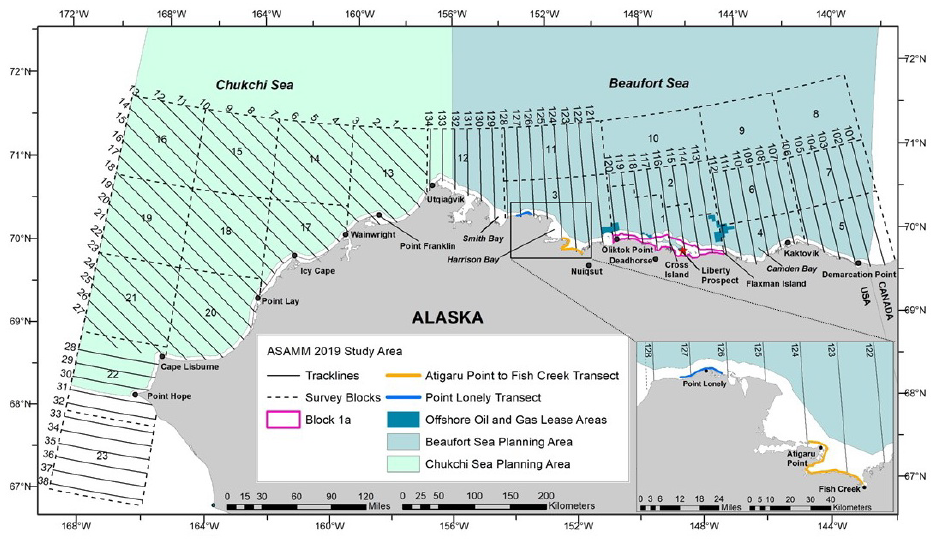}
    
    \caption{\emph{Aerial Surveys of Arctic Marine Mammals (ASAMM) 2019 study area spanning the northeastern Chukchi and central eastern Beaufort Seas. Solid lines show systematic aerial survey tracklines and dashed polygons indicate survey blocks. Colored shading marks federal offshore planning areas (Chukchi -green; Beaufort - blue) and associated lease blocks. Highlighted transects include Atigaru Point–Fish Creek (yellow), Point Lonely (blue), and Block 1a (magenta). Key coastal communities (e.g., Point Hope, Wainwright, Utqiaġvik, Nuiqsut, Kaktovik) provide geographic reference, and an inset shows a detailed view of the eastern Beaufort survey corridor.}}
    \vspace{0.5em}
    \label{fig:data_extent}
\end{figure}

After quality control, the dataset contained 10{,}592 geo referenced sightings from 236 flights, spanning $67.6^\circ$N--$72.9^\circ$N and $169^\circ$W--$118.9^\circ$W (Figure~\ref{fig:data_extent}). As part of the data pre processing step, we cleaned the coordinates by transforming all positions to EPSG:4326, checked for missing or duplicated timestamps, and screened against the Natural Earth coastline to remove or correct points falling on land. Observations lying exactly along land boundaries were approximated to the nearest valid ocean cell to avoid geometric inconsistencies during spatial modeling.

In the next step, temporal information was standardized to produce consistent year and month annotations. Because ASAMM survey effort is concentrated in the open water season, we restricted the analysis to July--October, thereby reducing temporal imbalance and excluding months lacking systematic coverage. The locations were later used to generate spatial exposure weights through Voronoi tessellation on the SPDE mesh and to ensure that the point process model accounted for variation in sampling intensity across flights, months, and years.

Environmental covariates were obtained by linking each sighting and mesh vertex to gridded sea surface temperature (SST) in Kelvin units from the Copernicus Climate Change Service (C3S) OISST product \citep{era5_single_levels}, as well as to distance to coast values computed from Natural Earth coastlines. Covariates were bilinearly interpolated to each observation and standardized (mean~0, sd~1) to improve model stability. To define the spatial domain, we constructed a high resolution coastline mask for the Chukchi Beaufort region and used it to constrain a Delaunay triangulation to generate the SPDE mesh. Mesh resolution was set to approximately 5~km in regions of dense observations and gradually coarsened offshore.
\begin{figure}[!htp]
    \centering
    \includegraphics[width=0.9\linewidth]{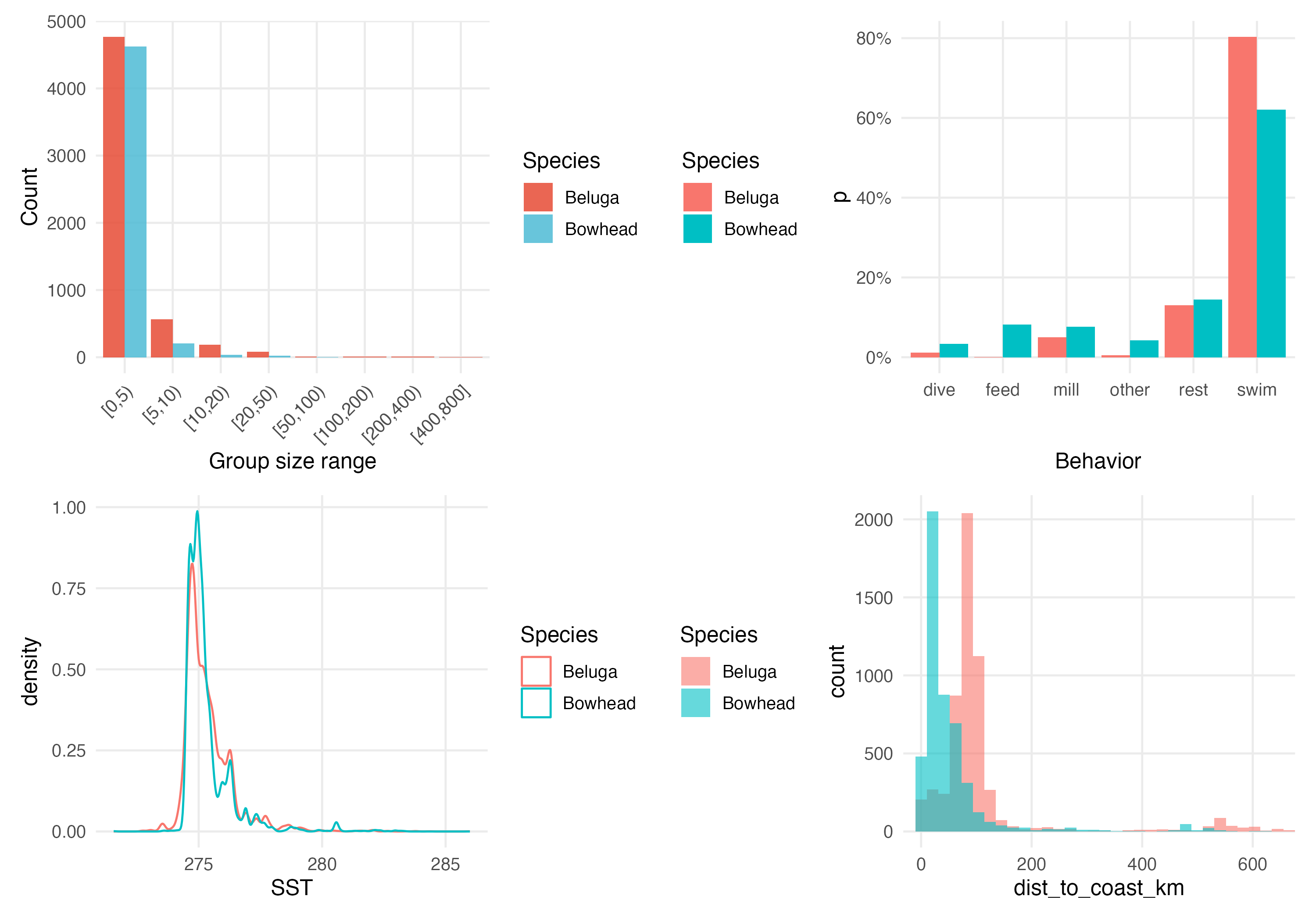}
    \caption{\emph{Exploratory summaries of ASAMM whale sightings (2010–2019), showing mostly small group sizes with occasional large beluga aggregations, dominant swimming and diving behaviors, strong coastal associations especially for belugas using shallow movement corridors \citep{Citta2020,Hauser2017} and similar cold water SST preferences for both species \citep{Hauser2017}}}\vspace{0.5em}
    \label{fig:Obs_dist}
\end{figure}
Figure \ref{fig:Obs_dist} compares the group size, behavior, and environmental context for beluga and bowhead whale sightings. Both species exhibit strongly right skewed group size distributions, with belugas more frequently occurring in small groups and bowheads occasionally forming larger aggregations. We also look at the behavioral composition of the two whale species. While swimming dominates for both species, bowheads show a higher share of feeding and milling behaviors. SST distributions indicate that both species occupy similarly cold surface waters with only subtle differences in thermal range. The distance to coast panel reveals a more nuanced habitat contrast. Bowheads (blue) are more common than belugas (red) within the first 0 - 40 km of the coast, likely reflecting their coastal migration corridor. Belugas dominate the 50 - 100 km zone, but then we also see both bowheads and belugas further off coast, which might be reflecting migration patterns. These patterns thus highlight distinct spatial niches and behavioral tendencies, that motivate species specific modeling in subsequent sections.

\begin{figure}[!htp] 
    \centering
    \includegraphics[width=.9\linewidth]{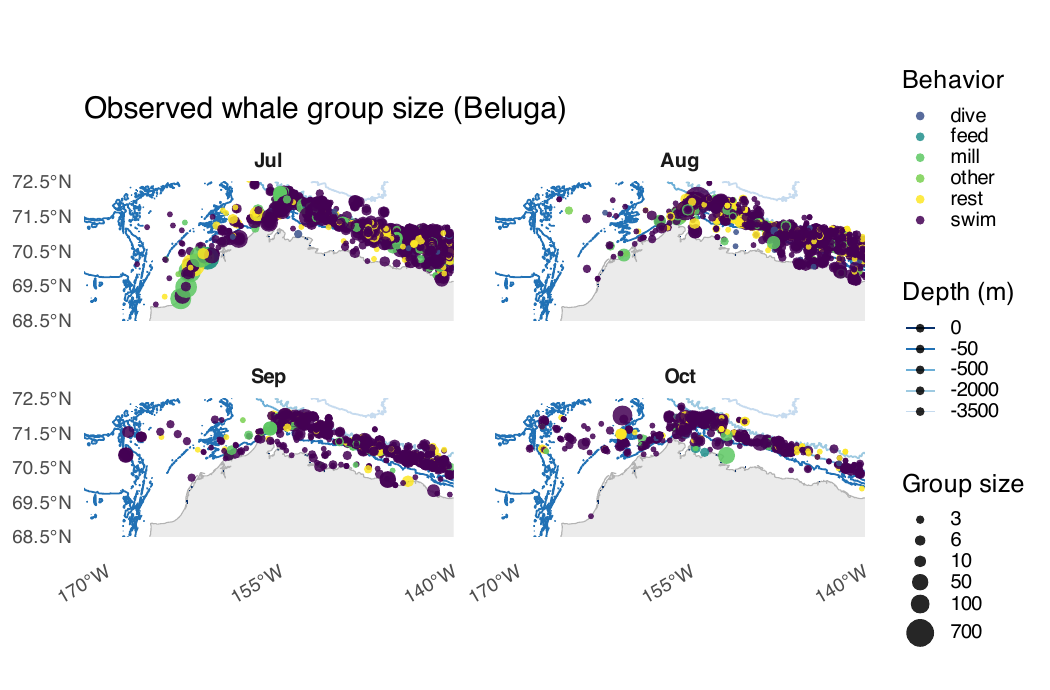}
    \caption{ \emph{Monthly distribution of observed beluga whale group sizes (July–October). Points show individual groups, with color indicating behavioral state and size proportional to reported group size. Bathymetric contours illustrate nearshore (dark) to offshore (light) depth gradients. Across months, belugas consistently use shallow coastal corridors in the northeastern Chukchi and central Beaufort Seas, with clear seasonal shifts in behavior and group size patterns.}} \vspace{0.5em}
    \label{fig:obs_beluga}
\end{figure}

\begin{figure}[!htp] 
    \centering
    \includegraphics[width=.9\linewidth,keepaspectratio=TRUE]{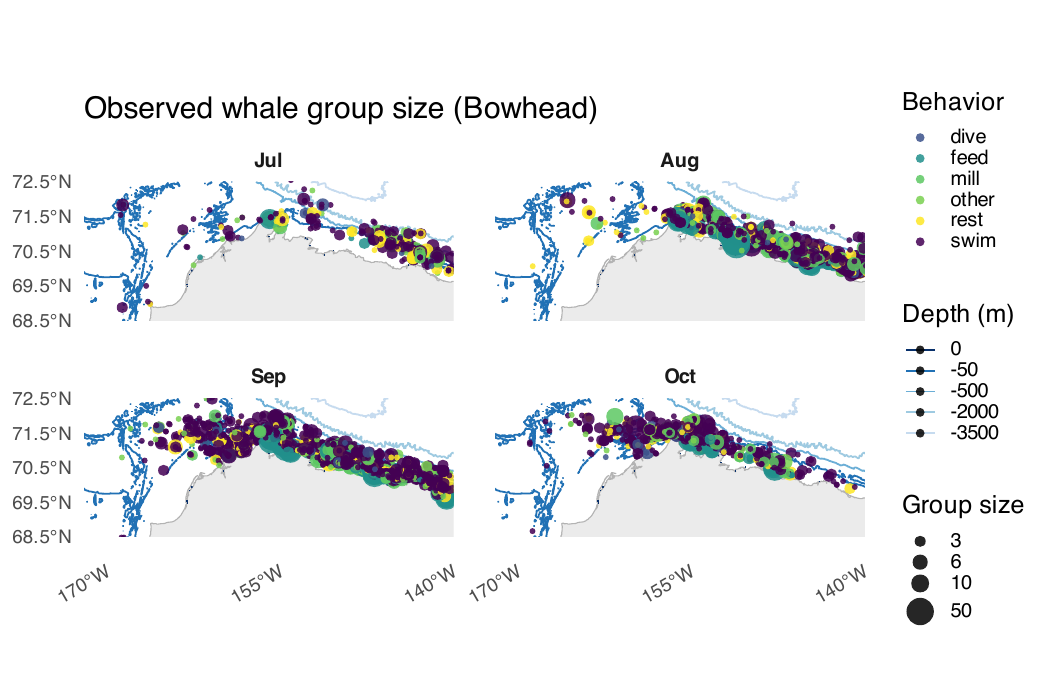}
    \caption{\emph{Monthly distribution of observed bowhead whale group sizes (July–October). Points represent individual groups, with color denoting behavioral state and size proportional to reported group size. Bathymetric contours highlight shallow coastal shelf areas (dark) and deeper offshore regions (light). The panels show seasonal movement along coastal migration corridors and shelf break habitats, with consistently nearshore summer distributions and clear spatial structuring of behavior and group size across the northeastern Chukchi and central Beaufort Seas.
} }\vspace{0.5em}
    \label{fig:obs_bow}
\end{figure}

Figure \ref{fig:obs_beluga} shows the seasonal distribution of beluga whale group sizes and behaviors across the Beaufort and Chukchi regions from July through October. Each monthly panel shows observed group locations for all years overlaid on bathymetric depth, with symbol sizes scaled by observed group sizes and color indicating behavioral state (namely, swim, dive, mill, feed, etc.). The plots highlight consistent nearshore use particularly along the Beaufort shelf and reveal clear seasonal shifts. Larger groups and more diverse behaviors appear in late summer (July–August), while sightings in September October become more dispersed and generally involve smaller groups. Similarly, figure \ref{fig:obs_bow} shows the seasonal distribution of bowhead whale group sizes and behaviors from July through October for all years, across the Beaufort and Chukchi Seas. Each monthly panel maps observed groups over bathymetric depth, with symbol sizes indicating group size bins and color denoting behavioral state. Bowheads occupy a broader offshore region than belugas, with large feeding and milling aggregations especially prominent in August and September along the Beaufort shelf and slope. By October, sightings shift westward and become more dispersed as whales migrate out of the region. 




\section{Joint Marked Log--Gaussian Cox Process (LGCP) Model}

We model whale group locations and their associated group sizes using a joint marked Log-Gaussian Cox Process (LGCP), in which the point pattern of sightings and the distribution of group sizes (marks) are linked through species specific latent Gaussian spatial fields. Let $g\in\{\mathrm{Beluga},\mathrm{Bowhead}\}$ index species, and let $\mathbf{s}\in\mathcal{D}\subset\mathbb{R}^2$ denote spatial location. The latent spatial field for species $g$ is denoted $w_g(\mathbf{s})$.

We represent each latent spatial field $w_g(\mathbf{s}), g\in \lbrace \mathrm{Beluga},\mathrm{Bowhead}\rbrace$, as a zero‐mean Gaussian Process (GP) with an anisotropic Mat\'ern covariance structure. Specifically, we assume
\[
w_g(\mathbf{s}) \sim \text{GP}(\bm{0}, \mathcal{C}(\kappa,\sigma^2,\nu,\mathbf{H})), \qquad \bm{s} \in \mathcal{D},
\]
where the covariance between two locations $\bm{s}, \bm{s}' \in \mathcal{D}$ is given by the anisotropic Mat\'ern form
$$
\mathcal{C}(\kappa,\sigma^2,\nu,\mathbf{H}) = \sigma^2 \frac{2^{1 - \nu}}{\Gamma(\nu)} \left( \kappa ||\mathbf{H}(\bm{s} - \bm{s}')||\right)^{\nu} K_{\nu}\left( \kappa ||\mathbf{H}(\bm{s} - \bm{s}')||\right).
$$
Here, $\sigma^2$ denotes the marginal variance, $\kappa > 0$ controls the spatial range, $\nu > 0$ is the Mat\'ern smoothness, $\mathbf{H} = \begin{pmatrix} h_x^2 & h_xh_yh_{xy} \\ h_xh_yh_{xh} & h_y^2\end{pmatrix}$ is a symmetric positive‐definite matrix encoding geometric anisotropy, and $K_{\nu}(\cdot)$ is the modified Bessel function of the second kind. The two species specific fields $w_{\mathrm{Beluga}}(\mathbf{s})$ and $w_{\mathrm{Bowhead}}(\mathbf{s})$ are treated as independent realizations of this Gaussian process but share the same hyperparameters. Following \citet{lindgren2011explicit, lindgren2022spde}, we implement the anisotropic Mat\'ern Gaussian random field through the stochastic partial differential equation (SPDE) representation, which connects the Mat\'ern family to a Gaussian Markov random field (GMRF) through an anisotropic Laplacian. In this framework, $w_g(\mathbf{s})$ is the stationary solution to
$$
\left( \kappa^2 - \nabla \cdot \mathbf{H}\nabla \right)^{\alpha/2} w_g(\mathbf{s}) = \frac{1}{\tau} \mathcal{W}(\bm{s}), 
$$
where, $\alpha = \nu + d/2$ ($d = 2$ in our application), $\tau > 0$ is a precision parameter related to the marginal variance by $\sigma^2 = \Gamma(\nu)/\lbrace \Gamma(\alpha)(4\pi)^{d/2}\kappa^{2\nu}\tau^2 \rbrace$, and $\mathcal{W}(\bm{s})$ is a spatial Gaussian white noise. We discretize this SPDE on a shared triangulated mesh covering the marine study region and use the INLA implementation to obtain computationally efficient GMRF approximations of the continuous Mat\'ern field. 

Now, conditional on $w_g(\mathbf{s})$, whale group locations are modeled as an inhomogeneous Poisson point process,
\[
N_g(\mathbf{s}) \mid w_g(\mathbf{s}) \sim \mathrm{Poisson}\{\lambda_g(\mathbf{s})\},
\]
with log–intensity
\[
\log \lambda_g(\mathbf{s})
= \alpha_{g} 
+ \beta_{g}\,d(\mathbf{s})
+ \gamma_{g}\,T(\mathbf{s};m,t)
+ \mathrm{Month}_{g,m}
+ \mathrm{Year}_{g,t} + w_g(\mathbf{s}),
\]
\[
\mathrm{Month}_{g,m} \sim \mathcal{N}(0,\tau^{-1}_{m,g}),
\qquad
\mathrm{Year}_{g,t} \sim \mathcal{N}(0,\tau^{-1}_{t,g}),
\]
where $\lambda_g(\mathbf{s})$ is the distribution intensity for species $g$ at location $\mathbf{s}$, 
$\alpha_g$ is a species specific intercept,
$d(\mathbf{s})$ denotes distance to coast, $T(\mathbf{s},m,t)$ is sea surface temperature at location $\bm{s}$ for month $m$ and year $t$, and  $w_g(\mathbf{s})$ is the latent Gaussian spatial field. 
The coefficients $\beta_g$ and $\gamma_g$ represent fixed effects describing the influence of distance and temperature, 
whereas $\mathrm{Month}_{g,m}$ and $\mathrm{Year}_{g,t}$ are species specific random effects for each month and year. 
These random effects differ from fixed coefficients in that they are modeled as zero mean Gaussian deviations whose variances 
$(\tau^{-1}_{m,g}$ and $\tau^{-1}_{t,g})$ are estimated from the data, allowing the model to capture unexplained seasonal and 
inter-annual variation.

Group size is treated as a mark attached to each observed location and is modeled conditionally on the same latent spatial field. For a detected group of species $g$ at location $\mathbf{s}$, the mark $Y_g(\mathbf{s})$ follows a negative binomial distribution,
\[
Y_g(\mathbf{s}) \mid w_g(\mathbf{s}) 
\sim \mathrm{NegBin}(\mu_g(\mathbf{s}), k_g),
\qquad
\log \mu_g(\mathbf{s})
= \alpha_g + \rho_g\,w_g(\mathbf{s})
+ \eta_{g}\,d(\mathbf{s})
+ \boldsymbol{\xi}_{g}^{\top}\mathbf{b}(\mathbf{s}),
\]
where $\mu_g(\mathbf{s})$ is the mean group size, $k_g$ is a species specific dispersion parameter, $\alpha_g$ is an intercept, $\rho_g$ determines the strength of the association between group size and the latent field, $\eta_{g}$ measures the effect of coastal proximity, and $\mathbf{b}(\mathbf{s})$ is a one hot vector encoding behavioral category at location $\bm{s}$ with coefficients $\boldsymbol{\xi}_{g}$. This hierarchical framework thus yields a coherent multi-species marked LGCP in which locations and group sizes depend on latent fields governed by a shared anisotropic Mat\'ern SPDE prior which ensures common spatial smoothness and directional dependence while still allowing each species to exhibit distinct spatial, environmental, and behavioral responses.


\subsection{Model Evaluation}\label{sec:eval}

We assess the predictive performance of the proposed model using the mean log score and the Watanabe Akaike Information Criterion (WAIC), computed on the scale of individual whale groups. The mean log score corresponds to the average log predictive density of the held out data, so higher (i.e., less negative) values indicate better predictive accuracy. The mean log score is computed as the average log posterior predictive density evaluated on the observed data,
\[
\overline{\mathrm{LS}}
= \frac{1}{n} \sum_{i=1}^{n}
\log\!\left( \int p(y_i\mid\theta)\, p(\theta\mid y)\, d\theta \right),
\]
which provides an in sample measure of predictive fit \citep{gneiting2007strictly}. Because the posterior is conditioned on the full dataset rather than leave one out quantities, this score represents the model’s internal predictive accuracy and complements the WAIC based assessment of out of sample performance.

WAIC is obtained from the joint posterior and reported per observation as $\text{WAIC} / n$. For completeness,  let $\ell_i(\theta) = \log p(y_i \mid \theta)$ denote the pointwise log likelihood contribution for observation $i$, where $\theta$ represents all model parameters and latent fields, and let $\mathrm{E}_{\mathrm{post}}[\cdot]$ denote posterior expectation. The log pointwise predictive density (lppd) is
\begin{equation*}
    \mathrm{lppd} = \sum_{i=1}^{n} 
    \log\!\left( \mathrm{E}_{\mathrm{post}} \left[ p(y_i \mid \theta) \right] \right).
\end{equation*}
The Watanabe Akaike Information Criterion is then defined as 
\begin{equation*}
    \mathrm{WAIC}
    = -2\bigl( \mathrm{lppd} - p_{\mathrm{WAIC}} \bigr), 
\end{equation*}
where $p_{\mathrm{WAIC}}$ denotes denote the effective numbers of parameters when calculating WAIC. Lower WAIC indicates a better balance between predictive accuracy and model complexity \citep{gneiting2007strictly}. For cross-species comparisons, we report WAIC per observation, $\mathrm{WAIC}/n$.

To assess spatial calibration of the fitted LGCP, we compute inhomogeneous $K$-function diagnostics \citep{baddeley2000non} and examine residual patterns to identify any departures from model assumptions. For a point process with spatially varying intensity $\lambda(\mathbf{s})$, the estimator is
\begin{equation*}
\widehat{K}_{\mathrm{inhom}}(r)
    = \frac{1}{|A|}
      \sum_{i \neq j}
      \frac{
        \mathbb{I}\{\lVert \mathbf{s}_i - \mathbf{s}_j \rVert \le r\}
      }{
        \widehat{\lambda}(\mathbf{s}_i)\, \widehat{\lambda}(\mathbf{s}_j)
      }
      w_{ij}^{-1},
\end{equation*}
where $\mathbf{s}_i$ and $\mathbf{s}_j$ are observed point locations, $|A|$ is the area of the observation window, $\widehat{\lambda}(\mathbf{s})$ is the fitted LGCP intensity, and $w_{ij}$ is the edge correction factor \citep{gneiting2007strictly}. If the empirical $\widehat{K}_{\mathrm{inhom}}(r)$ lies above the simulation envelope, the model underestimates clustering; if it lies below, it overestimates inhibition. We plot the normalized statistic 
$\widehat{K}_{\text{inhom}}(r) / (\pi r^2) - 1$, 
which removes the intrinsic $\pi r^2$ geometric growth of Ripley’s $K$--function 
\citep{ripley1976second,ripley1977modelling} 
and isolates spatial interaction beyond the modelled intensity surface.

\section{Results: Joint Spatio-temporal Model}\label{sec:results}
\subsection{Model Evaluation Results}
The joint marked model provides substantially stronger predictive performance for bowheads than for belugas. 
When evaluation is restricted to bowhead sightings (\(n = 4{,}669\)), the model attains a considerably higher mean log-score (\(-42.3\)) and a much lower WAIC per observation (\(83.1\)). 
In contrast, when evaluated on belugas (\(n = 5{,}624\)), the mean log-score decreases to \(-82.3\) and the WAIC per observation nearly doubles to \(164.0\). 
These differences indicate that the joint LGCP captures bowhead spatial intensity and group size structure far more effectively. 
Performance for belugas, while weaker, remains acceptable and reflects a modest trade off inherent to using an integrated multi species representation.
\begin{center}
\begin{minipage}{0.8\linewidth}
\centering
\captionof{table}{WAIC comparison for bowhead sightings: the joint marked LGCP far outperforms the baseline Poisson model.}
\label{tab:waic_base}
\begin{tabular}{lrr}
\toprule
\textbf{Model} & \textbf{WAIC} & \textbf{$\Delta$WAIC} \\
\midrule
Joint marked LGCP (bowhead) & $24{,}814.9$  & $0.0$ \\
Baseline Poisson            & $453{,}924.6$ & $429{,}109.7$ \\
\bottomrule
\end{tabular}
\end{minipage}
\end{center}

\begin{figure}[!htbp]
    \centering
    \includegraphics[width=0.8\linewidth]{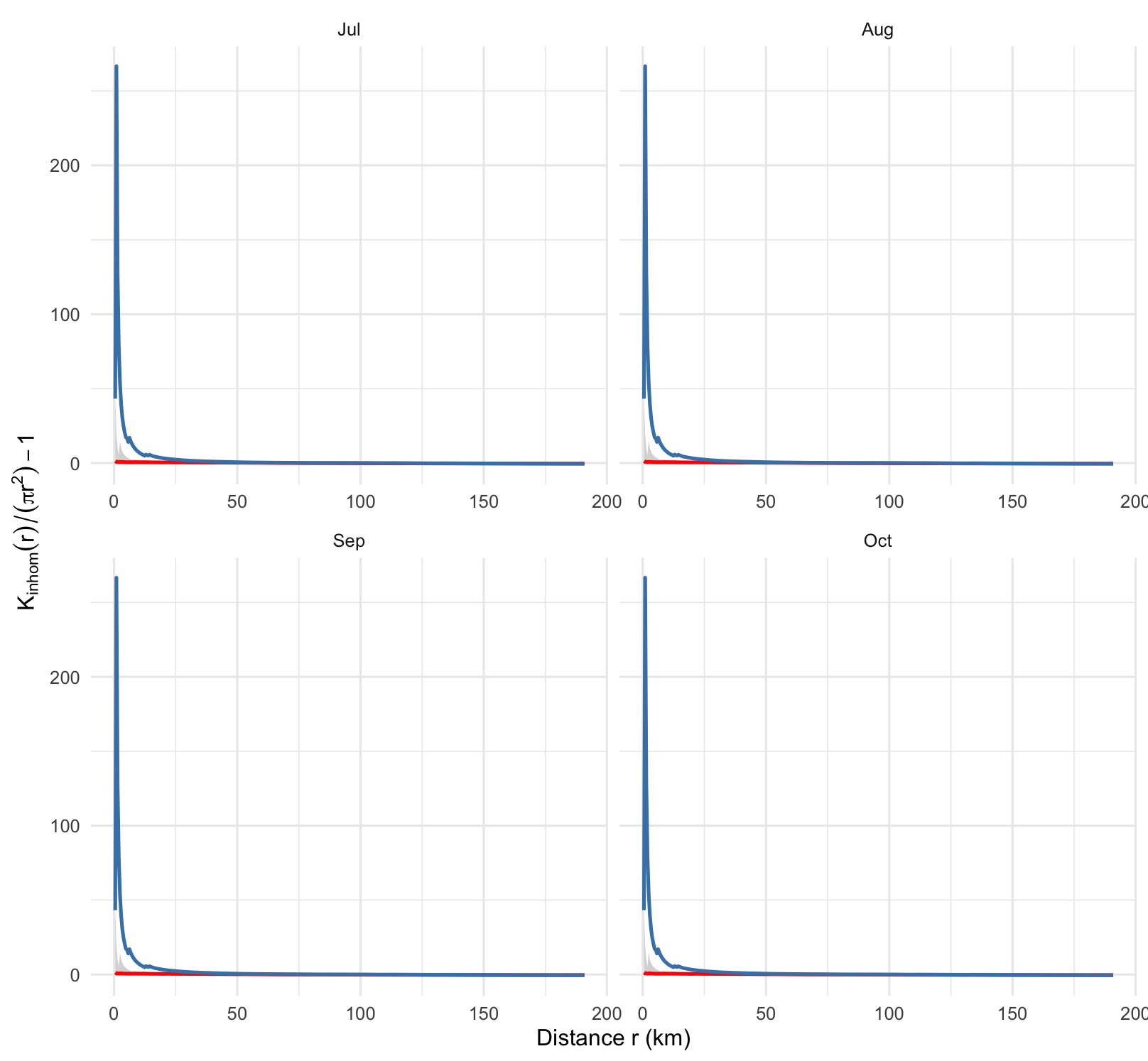}
        \caption{\emph{
Relative inhomogeneous $K$-function for evaluating second-order fit. 
The statistic $K_{\text{inhom}}(r)/(\pi r^{2}) - 1$ removes geometric $\pi r^{2}$ growth 
\citep{ripley1976second,ripley1977modelling} and isolates residual interaction. 
Empirical curves (blue) lie within 95\% simulation envelopes (gray) computed from 
99 posterior LGCP realizations \citep{baddeley2000non,baddeley2015spatial}, indicating good 
second--order fit with only slight small scale clustering ($<5$ km).
}} \vspace{0.5em}
    \label{fig:Kinhom_LGCP}
\end{figure}

\vspace{0.5em}
The joint marked LGCP yields an orders of magnitude reduction in WAIC compared with a baseline inhomogeneous Poisson process (IPP) model (Table \ref{tab:waic_base}), demonstrating a substantially better trade off between fit and effective complexity. Unlike the IPP, which models only the locations of whale sightings and assumes complete independence beyond the specified intensity surface, the LGCP incorporates latent spatial structure and jointly models both occurrence locations and group size marks. This leads to markedly improved predictive performance of the model achieved by capturing residual spatial dependence and incorporating biologically meaningful variation in group sizes. 
Relative inhomogeneous $K$–function diagnostics for the joint marked LGCP in Figure \ref{fig:Kinhom_LGCP} show the centered statistic $K_{\text{inhom}}(r)/(\pi r^{2}) - 1$ for each month (July–October), where dividing by $\pi r^{2}$ removes the geometric growth of Ripley’s $K$ and yields a zero reference value under an inhomogeneous Poisson process (\citep{ripley1976second,ripley1977modelling}) . Empirical curves (blue) are compared with 95\% simulation envelopes (gray) generated from 99 posterior LGCP simulations (\citep{baddeley2000non,baddeley2015spatial}). For all months empirical curves lie almost entirely within the envelopes, indicating that the fitted LGCP captures second order structure, with only mild excess clustering at distances $<5$ km.

\subsection{Model Results}
The proposed joint spatio-temporal Log-Gaussian Cox Process (LGCP) provides a quantitative framework for simultaneously estimating the spatial intensity of whale group occurrences, $\lambda_g(\bm{s})$, and the expected group size, $\mu_g(\bm{s})$ for the two whale species, $g \in \lbrace \mathrm{Beluga}, \mathrm{Bowhead}\rbrace$. The model converges successfully and yields stable posteriors for both components. Model diagnostics indicate strong overall fit, with a Deviance Information Criterion (DIC) of $-538147.1$, a Watanabe–Akaike Information Criterion (WAIC) of $101746.5$, and a marginal log-likelihood of $-333306.6$. The effective number of parameters based on WAIC was approximately $19667$ consistent with a moderately complex but well identified hierarchical model.
\begin{figure}[!htbp]
\centering

\includegraphics[width=\linewidth]{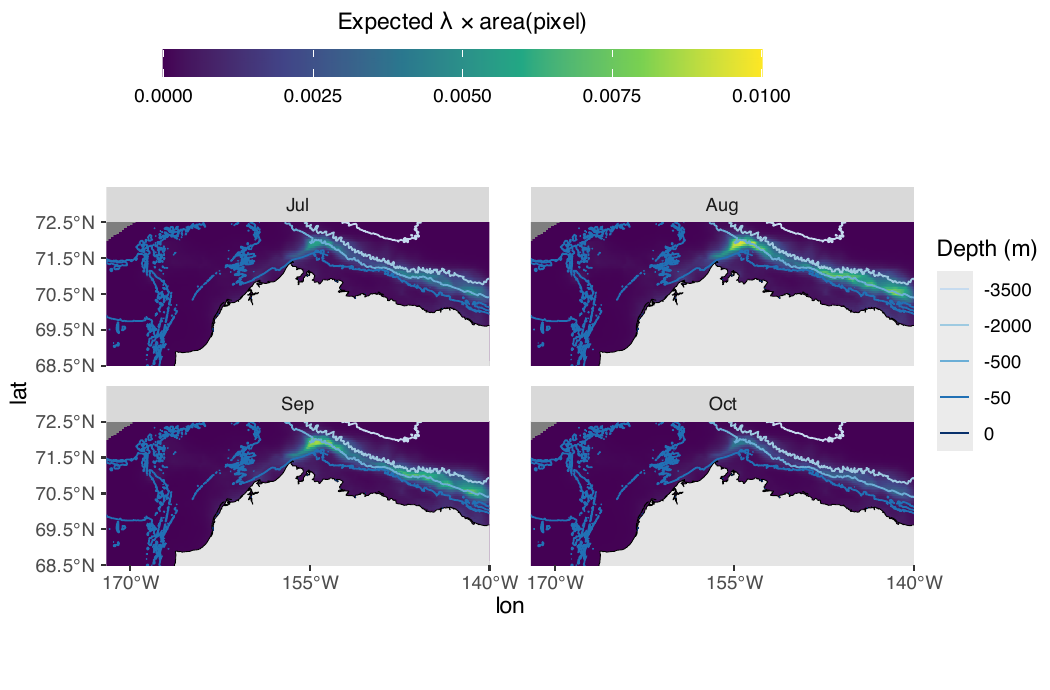}
    \caption{Predicted monthly spatial intensity $\lambda_g(\bm{s})$ (sightings per pixel) for beluga whales under the fitted joint LGCP model, evaluated on a 
regular prediction grid with pixel size of approximately 
4.5~km~$\times$~5~km. Each facet shows a month (July--October) using  a common color scale across months. Bathymetric depth contours are shown for reference.}\vspace{0.5em}
    \label{fig:intensity_beluga_sub}
\end{figure}

\begin{figure}[!htbp]
    \centering
    \includegraphics[width=\linewidth]{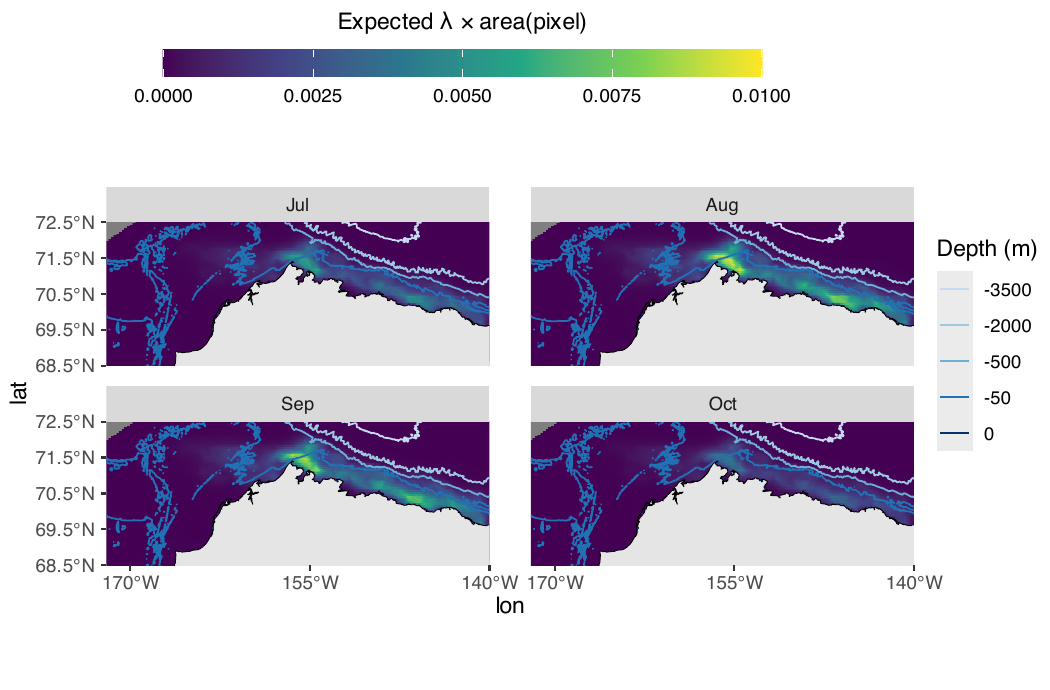}
    \caption{Predicted monthly spatial intensity $\lambda_g(\bm{s})$ (sightings per pixel) for bowhead whales under the fitted joint LGCP model. Intensities are evaluated on the same prediction grid used for belugas, corresponding to a pixel resolution of approximately $5 \times 5$~km across the study region. Each facet represents one month (July–October) and uses a common color scale to facilitate comparison across time. Bathymetric depth contours are overlaid to highlight the relationship between predicted intensity and shelf–slope structure.}\vspace{0.5em}
    \label{fig:intensity_bowhead_sub}
\end{figure}

At the process level, spatial predictions reveal clear and quantifiable differences in spatial intensities between species (Figures~\ref{fig:intensity_beluga_sub} and ~\ref{fig:intensity_bowhead_sub}). Mean posterior intensities $\lambda(\mathbf{s})$ for bowhead whales reach values on the order of $10^{-2}$ sightings per pixel, with highest densities occurring along the eastern Beaufort Sea shelf and shelf‐break. Peak intensities in August–September approach approximately $1\times 10^{-2}$, while most of the Beaufort Sea shows lower values ($<5\times 10^{-3}$).  
In contrast, spatial intensities for $g = \lbrace\mathrm{Beluga}\rbrace$ whales are lower and belugas are more spatially dispersed, with posterior means between $2.5\times10^{-3}$ and $7.5\times10^{-3}$ groups per pixel, which are concentrated in primarily nearshore Beaufort and Chukchi shelf waters. 
\begin{figure}[!htbp]
\centering

\includegraphics[width=\linewidth]{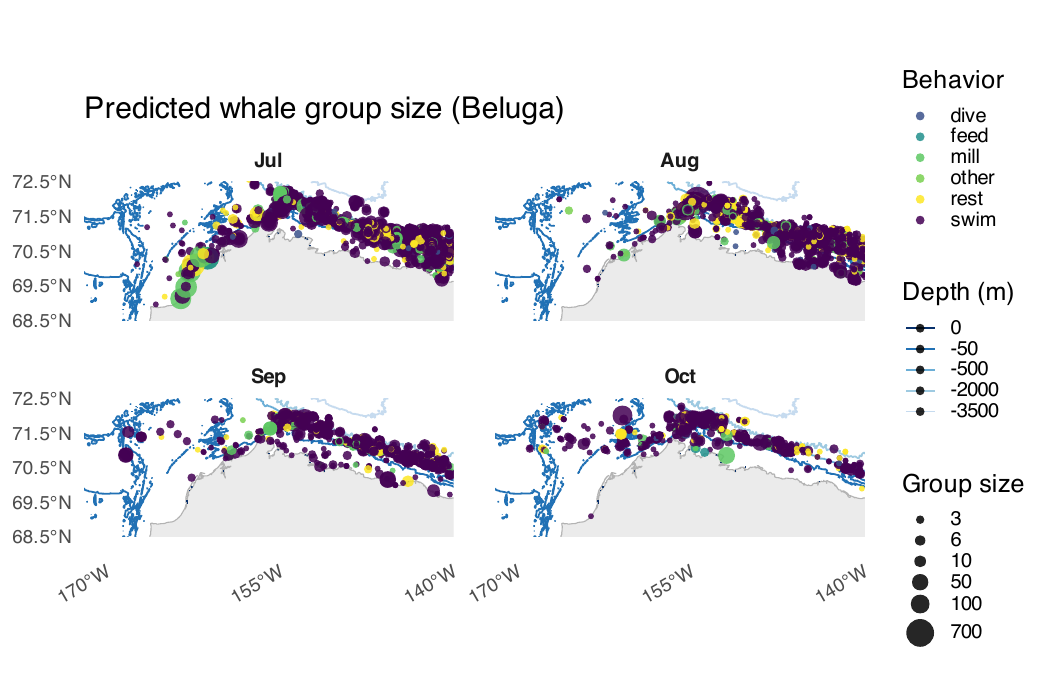}
    \caption{Predicted beluga whale group sizes from the marked LGCP model (July–October).
Points show predicted mean group sizes at observation locations, with size proportional to group size and color indicating behavioral state (dive, feed, mill, other, rest, swim). Bathymetric contours highlight shallow nearshore areas (dark blue). Beluga group sizes are largest in shallow coastal corridors during midsummer and early autumn, with seasonal shifts reflecting movement between the northeastern Chukchi and central Beaufort Seas.}\vspace{0.5em}
    \label{fig:mu_beluga_sub}
\end{figure}
Our results show that the mark component of the joint LGCP reveals strong and interpretable behavioral and environmental effects on whale group size for both species. For belugas, we see that behaviors exert substantial influence on predicted group size. Feeding exhibits the largest positive coefficient $(+4.24 \pm 0.44)$, followed by milling $(+2.32 \pm 0.05)$ and swimming $(+1.30 \pm 0.02)$, while resting $(+1.14 \pm 0.04)$ and diving $(+0.66 \pm 0.13)$ also increase expected group size of sightings. Distance to coast is negatively associated with group size $(-0.33 \pm 0.04)$, indicating that belugas form larger groups nearer to the shore. In contrast, coastal proximity is not a significant predictor for bowhead whale group sizes. Feeding $(+1.73 \pm 0.04)$ and milling $(+1.36 \pm 0.04)$ behaviors again emerge as dominant predictors of larger groups, while diving $(+0.83 \pm 0.06)$, resting $(+0.88 \pm 0.03)$, and swimming $(+0.87 \pm 0.02)$ show moderate positive contributions to group size. Bowhead group size shows strong positive effects of feeding and milling behavior but little dependence on coastal distance, whereas beluga group size increases sharply in nearshore, cooler, and feeding intensive environments. When viewed alongside the spatial intensity surface, these behavioral effects help differentiate the nearshore aggregation tendencies of belugas from the broader offshore patterns observed for bowheads.

Posterior summaries of the fixed effects for the spatial intensity process likewise show clear and ecologically coherent relationships (Table~\ref{tab:fixed_effects_full}). We see that the model intercept is estimated at $-26.0 \pm 0.4$, and distance to coast shows a strong negative effect $(-1.54 \pm 0.13)$, indicating reduced whale sighting probability offshore. Sea surface temperature (SST) also exhibits a modest negative association $(-0.13 \pm 0.034)$, suggesting slightly lower whale densities in warmer waters.

\begin{figure}
    \centering
    \includegraphics[width=\linewidth]{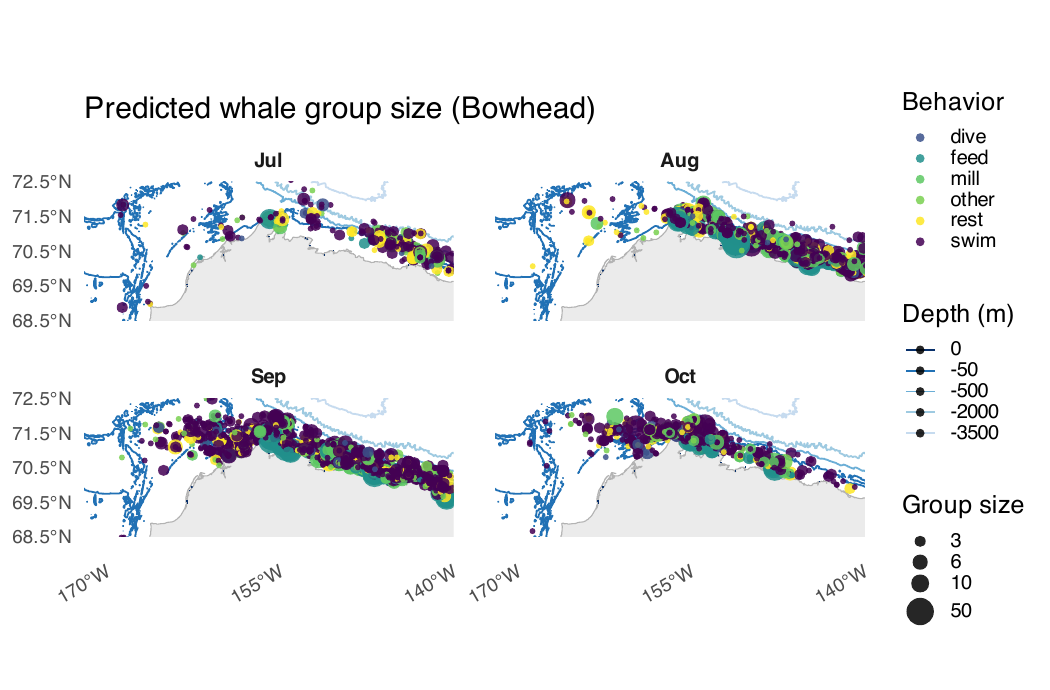}
    \caption{Predicted bowhead whale group sizes from the marked LGCP model (July–October).
Point size reflects modeled mean group size and color denotes behavioral state (dive, feed, mill, other, rest, swim). Bathymetric contours illustrate coastal shelf and offshore depth structure. Predicted group sizes are generally moderate but increase along the Beaufort inner shelf and coastal migration corridor, where feeding and milling behaviors coincide with larger aggregations. Seasonal patterns show a shift from the northeastern Chukchi toward the central Beaufort Sea from late summer into early autumn, consistent with known bowhead migration.} \vspace{0.5em}
    \label{fig:mu_bowhead_sub}
\end{figure}

Figures~\ref{fig:mu_beluga_sub} and \ref{fig:mu_bowhead_sub} illustrate how the fixed effects translate into spatially explicit predictions of whale group size across months. 
For belugas, predicted group sizes are largest in shallow nearshore waters from July to October, with notable shifts in location between the northeastern Chukchi and the central Beaufort Sea. 
For bowheads, predicted group sizes are more moderate overall but increase along the Beaufort inner shelf and shelf‐break where milling and feeding behaviors occur. 
Month‐to‐month changes in predicted group size distributions reflect the seasonal redistribution of both species across the Chukchi–Beaufort region.

\begin{table}[ht]
\centering
\caption{Posterior means and 95\% credible intervals for all fixed effects under the joint marked LGCP model. 
Coefficients correspond to the log linear scale of the intensity $\lambda_g(\bm{s})$ and expected group size $\mu_g(\bm{s})$ components, $g \in \lbrace \mathrm{Beluga}, \mathrm{Bowhead}\rbrace$.}
\label{tab:fixed_effects_full}
\begin{tabular}[H]{lccc}
\toprule
\textbf{Parameter} & \textbf{Mean} & \textbf{2.5\%} & \textbf{97.5\%} \\
\midrule
Intercept & $-26.000$ & $-26.951$ & $-25.271$ \\
Distance to coast ($s_{\text{dcoast}}$) & $-1.542$ & $-1.789$ & $-1.291$ \\
SST ($s_{\text{SST}}$) & $-0.130$ & $-0.197$ & $-0.063$ \\
Beluga: distance to coast & $-0.326$ & $-0.394$ & $-0.258$ \\
Beluga: dive & $+0.662$ & $+0.410$ & $+0.914$ \\
Beluga: feed & $+4.239$ & $+3.379$ & $+5.099$ \\
Beluga: mill & $+2.322$ & $+2.220$ & $+2.424$ \\
Beluga: other & $+0.778$ & $+0.412$ & $+1.145$ \\
Beluga: rest & $+1.136$ & $+1.066$ & $+1.206$ \\
Beluga: swim & $+1.304$ & $+1.274$ & $+1.334$ \\
Bowhead: distance to coast & $+0.036$ & $-0.023$ & $+0.094$ \\
Bowhead: dive & $+0.825$ & $+0.710$ & $+0.939$ \\
Bowhead: feed & $+1.732$ & $+1.664$ & $+1.800$ \\
Bowhead: mill & $+1.360$ & $+1.290$ & $+1.430$ \\
Bowhead: other & $+0.880$ & $+0.781$ & $+0.980$ \\
Bowhead: rest & $+0.883$ & $+0.821$ & $+0.945$ \\
Bowhead: swim & $+0.867$ & $+0.826$ & $+0.908$ \\
\bottomrule
\end{tabular}
\end{table}

Posterior summaries of the spatial hyperparameters (Table~\ref{tab:hyperparameters_joint}) show that 
the fitted model includes a well structured latent spatial field. 
INLA reports the four internal SPDE hyperparameters $\Theta_1$--$\Theta_4$ associated with the 
anisotropic Matérn representation, but these parameters are not directly interpretable as marginal 
variance or spatial range. To obtain meaningful covariance quantities, we transformed the posterior 
summaries of $\Theta_1$--$\Theta_4$ using 
\texttt{rSPDE::transform\_parameters\_anisotropic()}, which maps the internal SPDE parameters to the 
anisotropic Matérn parameters $(h_x, h_y, h_{xy}, \sigma)$ under the fixed smoothness $\nu = 2$. 
The resulting values (Table~\ref{tab:anisotropic_matern}) indicate a smoothly varying spatial field 
with moderate correlation ranges along the principal axes of anisotropy and a marginal standard 
deviation of approximately $\sigma \approx 1.4$.
\begin{table}[h]
\centering
\caption{Derived anisotropic Matérn hyperparameters for the shared spatial field in the joint marked LGCP, obtained by 
transforming the INLA SPDE hyperparameters $\Theta_1$--$\Theta_4$ using 
\texttt{rSPDE::transform\_parameters\_anisotropic}. 
Parameters $h_x$ and $h_y$ represent directional correlation scales along the principal axes of anisotropy, 
$h_{xy}$ governs the rotation and coupling between these axes, 
and $\sigma$ is the marginal standard deviation of the latent Gaussian field. 
Together, these values describe a moderately anisotropic spatial structure with correlation decaying over 
approximately 28--32 km along the dominant axes and a smoothly varying latent surface of amplitude $\sigma \approx 1.4$.}
\label{tab:anisotropic_matern}
\begin{tabular}{lccc}
\toprule
\textbf{Parameter} & \textbf{Mean} & \textbf{2.5\%} & \textbf{97.5\%} \\
\midrule
$h_x$ (km)   & 28.0 & 25.7 & 30.6 \\
$h_y$ (km)   & 32.2 & 29.5 & 35.2 \\
$h_{xy}$     & 0.42 & 0.32 & 0.51 \\
$\sigma$     & 1.41 & 1.32 & 1.52 \\
\bottomrule
\end{tabular}
\end{table}

Temporal precision parameters quantify the inverse of temporal variability in the random effects for year and month. 
The posterior mean precision for interannual variation ($3.74 \pm 1.65$) indicates moderate year to year differences in overall intensity, while the higher monthly precision ($11.95 \pm 7.90$) points to stronger, well resolved seasonal structure from July to October. 
Thus, the model captures consistent intra seasonal aggregation patterns, with relatively smaller long term shifts between years.

Finally, the negative binomial dispersion parameters reveal species specific differences in group size variability. 
Belugas exhibited greater overdispersion (size parameter $\approx 1.73$), implying a wider spread and more heterogeneous group sizes across space and behavior types, whereas Bowheads showed lower overdispersion (size parameter $\approx 16.3$), consistent with more stable, predictable group sizes. 
These contrasts indicate that Beluga group structure is more variable and responsive to local environmental and social factors, while Bowhead aggregations are more uniform across their core foraging areas.


\section{Discussion}\label{sec:discussion}
\begin{table}[htbp]
\centering
\caption{Posterior means and 95\% credible intervals for hyperparameters and random‐effect precisions in the joint marked LGCP. Spatial field hyperparameters ($\Theta_1$--$\Theta_4$) correspond to the internal SPDE parametrization of the anisotropic Matérn model; transformed covariance quantities are reported in Table~\ref{tab:anisotropic_matern}. Negative binomial size parameters ($1/\text{overdispersion}$) describe dispersion in group‐size marks.}
\label{tab:hyperparameters_joint}
\begin{tabularx}{\textwidth}{l *{3}{>{\centering\arraybackslash}X}}
\toprule
\textbf{Parameter} & \textbf{Mean} & \textbf{2.5\%} & \textbf{97.5\%} \\
\midrule
\multicolumn{4}{l}{\textit{Negative-binomial (group size) dispersion}} \\
Size (Beluga) & 1.727 & 1.652 & 1.805 \\
Size (Bowhead) & 16.293 & 13.986 & 18.937 \\
\addlinespace
\multicolumn{4}{l}{\textit{Spatial field hyperparameters}} \\
$\Theta_1$ & 10.241 & 10.153 & 10.329 \\
$\Theta_2$ & 10.381 & 10.294 & 10.469 \\
$\Theta_3$ & 0.896 & 0.673 & 1.122 \\
$\Theta_4$ & 0.346 & 0.275 & 0.417 \\
\addlinespace
\multicolumn{4}{l}{\textit{Temporal precision (random effects)}} \\
Year effect precision & 3.739 & 1.402 & 7.769 \\
Month effect precision & 11.952 & 2.547 & 32.267 \\
\addlinespace
\multicolumn{4}{l}{\textit{Copy field scaling (marks)}} \\
$\beta_{\text{field, Beluga}}$ & 0.025 & $-0.073$ & 0.123 \\
$\beta_{\text{field, Bowhead}}$ & 0.025 & $-0.073$ & 0.123 \\
\bottomrule
\end{tabularx}
\end{table}
This study provides a unified, quantitative view of where beluga and bowhead whale groups occur in the U.S. Arctic and how large those groups are. The joint marked LGCP model captures broad spatial heterogeneity in group occurrence 
(Figure~\ref{fig:intensity_beluga_sub} $\&$ \ref{fig:intensity_bowhead_sub}) as well as 
species specific differences in group sizes that depend on environmental and  behavioral factors (Table~\ref{tab:fixed_effects_full}). From the results, it can be seen that bowhead intensity peaks along the Beaufort shelf and coastal migration corridor, increasing from July to  September, while beluga intensity is lower overall and more fragmented, with  hotspots concentrated in nearshore channels and archipelago corridors. For belugas, 
expected group size increases closer to shore and during feeding and milling behaviors \citep{Hauser2017}, whereas bowhead group sizes are less tied to coastal proximity but tend to increase during feeding and milling activity \citep{Clarke2018}. These spatial and behavioral patterns are consistent with established species 
ecology. Belugas commonly form large aggregations in cold, nearshore environments, 
particularly in areas influenced by estuarine mixing, riverine outflow, and 
localized prey concentration. This behavior has been documented across multiple 
Arctic regions, with belugas selecting shallow coastal waters and estuarine 
habitats during summer where prey availability is elevated 
\citep{Hauser2017, Citta2018}. Bowhead whales, while often associated with 
shelf break fronts and offshore zooplankton aggregations later in the season, 
regularly use a coastal migration corridor in the eastern Beaufort Sea during 
summer and early fall. Satellite telemetry, aerial surveys, and passive acoustic 
monitoring show that bowheads travel and forage along the inner shelf and nearshore 
zone from July to September before shifting westward and offshore as autumn 
progresses \citep{clarke2020distribution, Clark2018Arctic, Fraker1998summer}. Our modeled 
intensity surfaces reflect this seasonal behavior, with bowhead occurrence peaking 
near the Beaufort coast and beluga hotspots concentrated in nearshore channels and 
archipelago corridors. Thus, the joint LGCP framework recovers the expected 
nearshore–offshore distinctions for each species during the July - October period 
and highlights behaviorally mediated aggregation patterns that align with 
previously documented habitat use.

Beyond spatial structure, the model also identifies important environmental and temporal effects. Sea surface temperature (SST) shows a modest but consistent negative association with whale occurrence intensity , indicating that the two species are more frequently observed in cooler waters, particularly during early and mid season months. This pattern aligns with established findings linking lower SST to increased prey availability and enhanced energetic efficiency in Arctic feeding grounds. Temporal random effects reveal clear seasonality from July through October, with posterior means showing a progressive increase in overall intensity through late summer and early autumn, corresponding to peak aggregation and foraging periods. Yearly random effects capture moderate interannual variation, as indicated by the posterior precision estimate. 
Such year‐to‐year differences are consistent with documented interannual variability in environmental conditions affecting bowhead and beluga habitat use \citep[e.g.,][]{Ashjian2010,Druckenmiller2018,Stafford2021}, 
as well as known differences in survey coverage across years in the ASAMM program \citep[e.g.,][]{Clarke2018}.

The inferred coupling between marks and intensity is weak (Table~\ref{tab:hyperparameters_joint}), suggesting that whale occurrence and group size are governed by partially distinct spatial processes. While first order occurrence patterns likely reflect broad habitat suitability and survey availability, group size appears to respond to more localized ecological and social dynamics. For beluga whales, large aggregations are often associated with known calving and molting areas where individuals congregate seasonally for reproductive or physiological purposes. The eastern Chukchi Sea stock, for example, calves and molts along the northwestern coast of Alaska, particularly offshore of and within Kasegaluk Lagoon, in late June through early July \citep{Suydam2001, DeMaster1998}. The Beaufort Sea stock exhibits similar behavior in the Mackenzie River Delta, Canada, where large summer concentrations have been consistently documented \citep{DFO2005, Mayette2023}. Substantial groups are observed offshore of both regions by ASAMM and other studies, and our fitted model likewise identifies these large aggregations in the same areas, indicating that localized biological activity rather than general habitat preference can drive exceptionally large beluga group sizes. Accordingly, the weak spatial correlation between intensity and group size likely reflects the influence of these concentrated biological and seasonal behaviors operating within broader distributional patterns.

While the joint marked LGCP provides a comprehensive framework for modeling whale distributions, several caveats warrant consideration. Although the model conditions on observed sightings and accounts for temporal variation through random effects, unmodeled heterogeneity in survey effort, observer conditions, and distance‐related detectability may still influence the estimated intensity and mark processes. Incorporating explicit effort offsets or distance sampling corrections within the LGCP structure would help mitigate such detectability biases. The environmental covariates used in this study are limited. Additional oceanographic and biological processes, such as sea ice concentration, frontal activity, chlorophyll concentration, bathymetric complexity, and prey availability could explain additional spatial or temporal variability in the data. Temporal alignment between covariates and observation timing (e.g., monthly SST versus daily survey effort) may also smooth finer scale ecological responses. Spatially, the model uses an anisotropic Mat\'ern covariance to capture directional dependence along the Beaufort shelf and coastal channels. However, residual nonstationarity may persist in complex nearshore regions where environmental gradients and movement pathways interact. Moreover, because the analysis focuses on the July–October survey season, the estimated spatial range and variance parameters primarily reflect summer distributions and may not extrapolate directly to other seasons.


In conclusion, the joint marked LGCP framework provides a powerful and flexible basis for understanding and managing whale–vessel interactions in the United States Arctic. The spatial intensity maps, with associated uncertainty, identify high use whale corridors such as the Beaufort shelf for bowheads and nearshore channels for belugas, and they reveal how these patterns evolve from July through October. These predictions can directly inform dynamic vessel routing strategies, temporal speed reductions, and seasonal area management measures that balance maritime operations with ecological conservation goals. The predicted group sizes derived from the mark component translate the spatial surfaces into expected numbers of individuals per aggregation, providing a quantitative foundation for estimating exposure overlap when combined with Automatic Identification System (AIS) based vessel density data. Because the correlation between the mark and intensity processes is weak, management strategies that focus solely on high occurrence areas may overlook locations where fewer but substantially larger groups are present. Both spatial intensity and group size patterns should therefore guide mitigation planning, including the placement of seasonal slow zones, adjustments to vessel routing, and the scheduling of high traffic industrial activities such as shipping, oil and gas operations, and mining transport.
\section*{Data Availability Statement}

All whale sighting data used in this study were obtained from publicly 
available Aerial Surveys of Arctic Marine Mammals (ASAMM) annual reports 
published by NOAA. Reports from 2010–2019, including raw sighting tables 
provided in their appendices, are accessible through the NOAA Fisheries 
website at 
\url{https://www.fisheries.noaa.gov/alaska/marine-mammal-protection/aerial-surveys-arctic-marine-mammals}.
These data are openly available and do not require special access 
permissions.

Sea-surface temperature (SST) covariates were obtained from the ERA5 
reanalysis produced by the European Centre for Medium-Range Weather 
Forecasts (ECMWF). ERA5 hourly SST fields on single levels were accessed 
through the Copernicus Climate Data Store 
(\url{https://cds.climate.copernicus.eu/datasets/reanalysis-era5-single-levels}). 
The downloaded data were provided in GRIB format and converted to 
CF-compliant NetCDF using the \texttt{cfgrib} interface.

R codes used to implement the proposed model will be made public on GitHub upon publication of the paper.
\section*{Funding}
This work received no external funding and was conducted as part of the 
author’s doctoral research in Integrative Life Sciences Doctoral Program, Center for Integrative Life Sciences Education, 
Virginia Commonwealth University.
\section*{Conflict of Interest}
The author declares no conflicts of interest.
\bibliography{wileyNJD-APA}
\end{document}